\documentclass[a4paper,12pt]{article}
\usepackage[affil-it]{authblk}
\usepackage{graphics}
\usepackage{color}
\usepackage{mathtools}
\usepackage{amsmath}
\usepackage{subcaption}
\usepackage{mathtools}
\usepackage[normalem]{ulem}
\usepackage{float}
\usepackage{stfloats}
\usepackage{balance}
\usepackage{lineno,hyperref}
\usepackage{cite}

\usepackage{fullpage}
\begin{document}

\title{Reentrant condensation transition  in a  two species driven diffusive system}
\author{Bijoy Daga
\thanks{Electronic address: \texttt{bijoy.daga@saha.ac.in} }}
\affil{CMP Division, Saha Institute of
Nuclear Physics, HBNI, 1/AF Bidhannagar, Kolkata 700064, India}

\date{}

\maketitle






\begin{abstract}
We study an interacting box-particle system on a one-dimensional periodic ring involving two species of particles $A$ and $B$. In this 
model, from a randomly chosen site, a particle of species $A$ can hop to its right neighbor with a rate that depends on
the number of particles of the species $B$ at that site. On the other hand, particles of species $B$ can be transferred between 
two neighboring sites with rates that depends on the number of particles of species $B$ at the two adjacent sites$-$this process 
however can occur only when the two sites are devoid
of particles of the  species $A$. We study condensation transition for a specific choice of rates and find that 
the system shows a reentrant phase transition of species $A$ $-$ the species $A$ 
passes successively through fluid-condensate-fluid phases as the coupling parameter between the dynamics of the two species is varied. On the other hand,
the transition of species $B$ is from condensate to fluid phase and hence does not show reentrant feature.
\end{abstract}



%
%

\section{Introduction}
\label{sec:intro}
The study of systems driven far from thermal equilibrium, also known as driven diffusive systems (DDS), has been at the forefront in statistical 
mechanics in the last few decades  \cite{chowdhury_book}. 
These systems have found applications in understanding
transport in superionic conductors \cite{KLS1, KLS2}, protein synthesis 
in prokaryotic cells \cite{gibbs_protein1, gibbs_protein2}, 
traffic flow \cite{traffic}, biophysical transport \cite{chou_bio_transport,frey_bio_transport}, etc. These systems
evolve under local stochastic dynamics and in the long time limit reach a non-equilibrium current carrying stationary state.
Certain surprising features of these non-equilibrium steady states have generated an overwhelming interest among researchers. 
For example, these systems may exhibit spontaneous symmetry breaking \cite{ssb_dds}, boundary induced 
phase transition \cite{krug_asep, popkov_schutz}, phase separation transition\cite{ABC_model1}, condensation 
transition \cite{evans_braz, zrp_review} even 
in one dimension. 

\par In this work, we discuss reentrant condensation transition in a DDS involving two species 
of interacting particles on a one dimensional periodic ring. A reentrant phase transition is said to occur 
if by varying a certain parameter, 
the system undergoes transition from one phase to another phase and finally reenters the initial phase. Such transitions have been 
reported in a variety of equilibrium systems, for example  in models of spin glasses \cite{reentrance_spin_glass} and multicomponent 
liquid mixtures \cite{reentrance_liquid_mix}. Reentrant transition have also been reported in some non-equilibrium systems. 
For example, Antal and Sch\"utz studied a system 
of a driven non-equilibrium  lattice gas of hard-core particles with next-nearest neighbor interaction in one
dimension \cite{schutz}, where 
for attractive interactions, a reentrant transition between high density (HD) phase and maximal current(MC) 
phase was observed. A few reaction-diffusion systems \cite{dickman_pcpd,odor_pcpd1} having  a 
competing dynamics between 
diffusion and particle interaction have been observed to show a similar reentrant phase behavior$-$ where 
the transition is from absorbing-to-active-to-absorbing  phases.
In  biological systems, reentrant transitions  have been reported experimentally in protein and DNA solutions in presence of 
multivalent metal ions \cite{zhang_protein_reentrance}. 
Such physical phenomena are vital for understanding biological processes like 
protein crystallization and DNA condensation. Reentrant phase behavior has also been observed in driven colloidal
systems \cite{hagan_colloids2} and force induced DNA unzipping transitions \cite{bhattacharjee_DNA_unzipping}.

\par The two species driven diffusive model discussed here was first introduced in \cite{our_model}, where the phase separation transition 
in a model of reconstituting $k$-mers can be studied by mapping the model to the two species box particle system. Here
we report that this  model can  show a reentrant  condensation transition where one of the species undergoes fluid-condensate-fluid transition 
when the interaction parameter between the two species is varied. 

\par The article is organized  in the following way: In section 2 we define the model and  write down its product measure steady state. 
In the next section we study the  system   in the grand canonical ensemble for a specific choice of diffusion rates 
and show that condensation transition in one of the species show a reentrant behavior. In section 4  we  compare the reentrant 
behavior of this model with that in some single species models and finally we summarize the results in section 5 and conclude
with some discussions.

\begin{figure}
\centering \includegraphics[height=5cm]{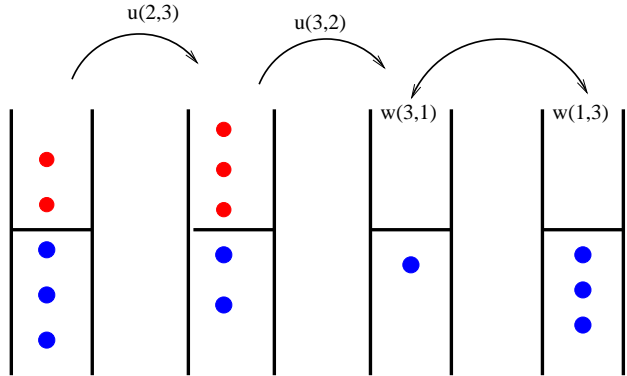}
\caption{ Dynamics of the two-species box-particle model: The diffusion rate, $u(m,n)$ of species $A$ (red-circles) in general can 
depend on the 
number of particles
of both the species  at that site. Exchange process of species $B$ (blue-circles) can occur only if both the  
sites are devoid of the species $A$.
The rate of this process depends on the number of particles of the species $B$ at the two participating sites. 
}
\label{reentrant2}
\end{figure}

\section{The model }
\label{sec:model}
We study condensation transition in a  box-particle model on a one dimensional periodic lattice of $L$  sites or boxes
labeled  as $i=1,2,\cdots,L$
and  containing two species of  particles, say $A$ and $B$. At a given site $i$, let $m_i$ and $n_i$ be the 
number of particles of  species $A$ and $B$ respectively. A typical configuration of the model is represented as  
$\ {C}= \{ m_1,n_1;m_2,n_2 \cdots m_i,n_i\cdots m_L,n_L\}=\{ m_i,n_i\}$. The dynamics of the model 
is the following: from a randomly chosen  site $i$, a particle of species $A$ can hop to site $i+1$ with rate $u(m_i,n_i)$ that 
in general depends on the number of particles of
both  $A$ and $B$. On the other hand, particles of species $B$  can be transferred  between two neighboring boxes 
$i$ and $i+1$ with rates that depend on the number of particles at both arrival and departure sites. However, in this case 
there is  an additional  restriction  that the boxes exchanging particles of species $B$ are devoid of 
particles of the  species $A$. This condition is crucial for explicit factorization 
of the steady state \cite{our_model}. Depending on whether a $B$-type particle is 
transferred from site $i$ to $i+1$ or vice-versa, we define $w(n_i, n_{i+1})$ or  $w(n_{i+1}, n_{i})$  
to be the corresponding rates for the dynamics of species $B$. The above dynamics, also
shown in Figure~\ref{reentrant2} can be 
represented in the following way:

\begin{equation}
\{ \cdots m_i,n_i;~m_{i+1},n_{i+1}\cdots\} \xrightarrow{u(m_i,n_i)} \{\cdots m_i-1,n_i;~m_{i+1}+1,n_{i+1}\cdots \},
\label{dyn1}
\end{equation}
\begin{equation}
\{\cdots 0,n_i;~0,n_{i+1} \cdots\} \xrightarrow{w(n_i,n_{i+1})} \{\cdots 0,n_i-1;~0,n_{i+1}+1 \cdots\}
\label{dyn2}
\end{equation}
and
\begin{equation}
\{\cdots 0,n_i;~0,n_{i+1}\cdots\} \xrightarrow{w(n_{i+1},n_i)} \{\cdots 0,n_i +1;~0,n_{i+1}-1 \cdots\}.
\label{dyn3}
\end{equation}

Note that the  dynamics conserves the total number of particles of each species.
In this article we would limit  to the 
situation where $u(m,n) \equiv u(n)$, thus the rate of diffusion of species $A$ 
depends on the number of particles of species $B$. Such a choice 
makes the dynamics of the species $A$ comparable to that of a disordered zero range process \cite{inhomo_zrp}, 
the background disorder being created by the presence of species $B$ evolve with time. A two 
species zero-range process \cite{2sp_zrp1, gross_spohn_03, 2sp_zrp2} has been 
previously studied where one of the species obeys a dynamics similar to that of species $A$ in this model. The dynamics of the species $B$ is similar 
to that of a misanthrope process \cite{misanthrope_original, misanthrope} and hence
the rate of diffusion depends on the number of particles in both departure and the arrival site. It is necessary to mention here that 
in the model of reconstituting $k$-mers, particles of species $A$ corresponds to $0$-particles 
and that of species $B$ corresponds to $k$-particles \cite{our_model}. 
\par The steady state of the model has a product measure so that the probability 
of finding the system in an arbitrary configuration, $\ {C}= \{ m_i,n_i \}$ 
can be written in  the following factorised form:
\begin{equation}
P(\{m_i,n_i\}) = \frac{1}{Q_{M,N}^L} \prod_{i=1}^L f(m_i,n_i) \delta\left(\sum_{i=1}^L m_i-M\right)\delta\left(\sum_{i=1}^L n_i-N\right),
\label{steady_state}
\end{equation}
where the partition function $Q_{M,N}^L$ is given by 

\begin{equation}
Q_{M,N}^L = \sum_{\{ m_i\}, \{n_i\}}\prod_{i=1}^Lf(m_i,n_i)
\delta\left(\sum_{i=1}^L m_i-M\right)\delta\left(\sum_{i=1}^L n_i-N\right).
\label{partition_func}
\end{equation}

The $\delta$-functions appearing above make sure that 
only those configurations are to be summed over which 
the number of particles $M$ and $N$ of species $A$ and $B$ are conserved. Therefore 
the densities, $\rho_A=M/L$ and $\rho_B=N/L$ of the two species  $A$ and $B$ do not change as 
the system evolves over time. Here, we consider factorized rate for species $B$, 
\begin{equation}
 w(m,n)= w_1(m) w_2(n).
 \label{rate1}
\end{equation}
This choice makes the steady 
state factorised and the weight factors appearing in Eq.~\eqref{steady_state} are then given by \cite{our_model}

\begin{equation}
f(m,n) = u(n)^{-m} \prod_{i=1}^n\frac{w_2(i-1)}{w_1(i)}.
\label{prod_measure}
\end{equation}
In the next section, we study the nature of condensation transition of the system for certain specific choice of rates.

\section{Condensation and reentrant transition}
\label{sec:cond_transition}
A salient feature of box-particle systems is that they can show condensation transition \cite{zrp_review, evans_pair, misanthrope}. For such systems, 
condensation occurs when the 
density $\rho$ is larger than a finite critical density $\rho_c$.  In the condensed phase there exists 
a background critical fluid consisting of $\rho_c L$ particles,
and a condensate carrying $(\rho -\rho_c)L$ particles, $L$ being the system size. For certain choices of diffusion rates, the two 
species model discussed here 
also shows condensation of one or both the species which can be established by studying the system in the grand 
canonical ensemble (GCE) \cite{our_model}. In GCE, let $z$ and $x$ be the fugacities corresponding to the species $A$ and $B$ respectively; then
the grand canonical partition function, $Z_L(z,x)$  obtained from the 
canonical partition function (Eq.~\eqref{partition_func}) is given by
\begin{equation}
Z_L(z,x)= \sum_{M=0}^{\infty} \sum_{N=0}^{\infty} z^M x^N Q_{M,N}^L= F(z,x)^L,
\label{grand_part_func}
\end{equation}
where
\begin{equation}
F(z,x)= \sum_{m=0}^{\infty} \sum_{n=0}^{\infty} z^m x^n f(m,n).
\label{grand_f_func}
\end{equation}
Let $z \le z_c$ and $x \le x_c$ define the domain of fugacities for which $F(z,x)<\infty$ and thus the grand canonical measure is valid within this domain. 
In GCE, densities are given by
\begin{eqnarray}
\rho_A(z,x)=\frac{z}{F}\frac{\partial F}{\partial z} ~~ {\rm and}~~
\rho_B(z,x)=\frac{x}{F}\frac{\partial F}{\partial x}~.
\label{grand_eq_den}
\end{eqnarray}
If one or both densities become finite 
as critical values of fugacities are approached then the system is in a condensate carrying phase. In such a  phase a system cannot
accommodate densities that are larger than the ones set by the critical limit and therefore undergoes a condensation transition.
Here, we discuss one of the rates for which the condensation transition of species $A$ shows a reentrant behavior.
Let us set the rate of diffusion of the species $A$ as a step function,

\begin{equation}
 u(n) = \left\{\begin{array}{ll}
          v ~~~&{\rm for  }~ n < k\cr
          1~~~ &{\rm otherwise}~.
         \end{array}
\right.
\label{diff_A_species}
\end{equation}
Here $k<\infty$ can take only positive integer values and $u(0)=0$. We also consider $w_1$ and $w_2$ to differ by a constant \cite{misanthrope}, 
\begin{equation}
w_1(n)=  \frac{n+2}{n+1}  ~~;~~w_2(n)= \frac{n+2}{n+1}-\alpha, 
\label{diff_B_species2}
\end{equation}
where $0<\alpha<1$. Defining $\sigma= (3-2 \alpha)/(1-\alpha)$, we characterise the rates of diffusion of 
species $A$ and $B$ by three  parameters, namely $v$, $k$ and $\sigma$. Using Eq.~\eqref{prod_measure}, we then obtain
\begin{eqnarray}
 f(m,n)=\frac{1}{[u(m)]^n}\frac{n!}{(\sigma)_n}(n+1)^2,
\label{eq:fk0}
\end{eqnarray}
where $(\sigma)_n=\sigma(\sigma+1)\cdots(\sigma+n-1)$ is the Pochhammer symbol. 
The partition function in the GCE, following Eqs. 
\eqref{grand_part_func} and \eqref{grand_f_func}, is $Z_L(z,x)= F(z,x)^L$, with
\begin{equation}
F(z,x)= \frac{z(1-v) G_{k -1 }(x) + (v-z) G_{\infty}(x)}{(1-z)(v-z)}, 
\label{f_func_choice}
\end{equation}
where
\begin{equation}
G_{k}(x)=\sum_{n=0}^{k} x^n \frac{n!}{(\sigma)_n}(n+1)^2.
\end{equation}
From Eq.~\eqref{f_func_choice}, it is evident that the maximum value of the 
fugacities for which $F(z,x)<\infty$  are $z_c=$min$ \{1,v\} $ and $x_c=1$. The  densities of the two species 
in GCE obtained from Eq.~\eqref{grand_eq_den} are given by

\begin{eqnarray} 
\rho_A(z,x)&=& \frac{z}{1-z} + \frac{ z v(1-v)G_{k-1}(x)}{(v-z)[z(1-v) G_{k-1}(x) + (v-z) G_\infty(x)] }\cr
\nonumber \\
{\rm and~}\nonumber\\
\rho_B(z,x) &=& x \frac{z( 1-v) G_{k-1}'(x) + (v-z) G'_\infty(x)}{z ( 1-v) G_{k-1}(x) + (v-z) G_\infty(x)}\nonumber\\.
\label{eq:densities}
\end{eqnarray}

Here, prime ($'$) indicates derivative with respect to $x.$ It has been shown in \cite{our_model} that as $x \to x_c (=1)$,  $\rho_B$ 
may become finite but as $z \to z_c$, $\rho_A$ always diverge. This means that in absence 
of $A$-type particles, $i.e.$ when $z=0$, species $B$ can condense but $A$ cannot condense if there are no particles of species $B (x=0)$  in the 
system. Therefore
the  critical line  is given by  $x=1$.

\begin{figure}
\centering \includegraphics[height=5.5cm]{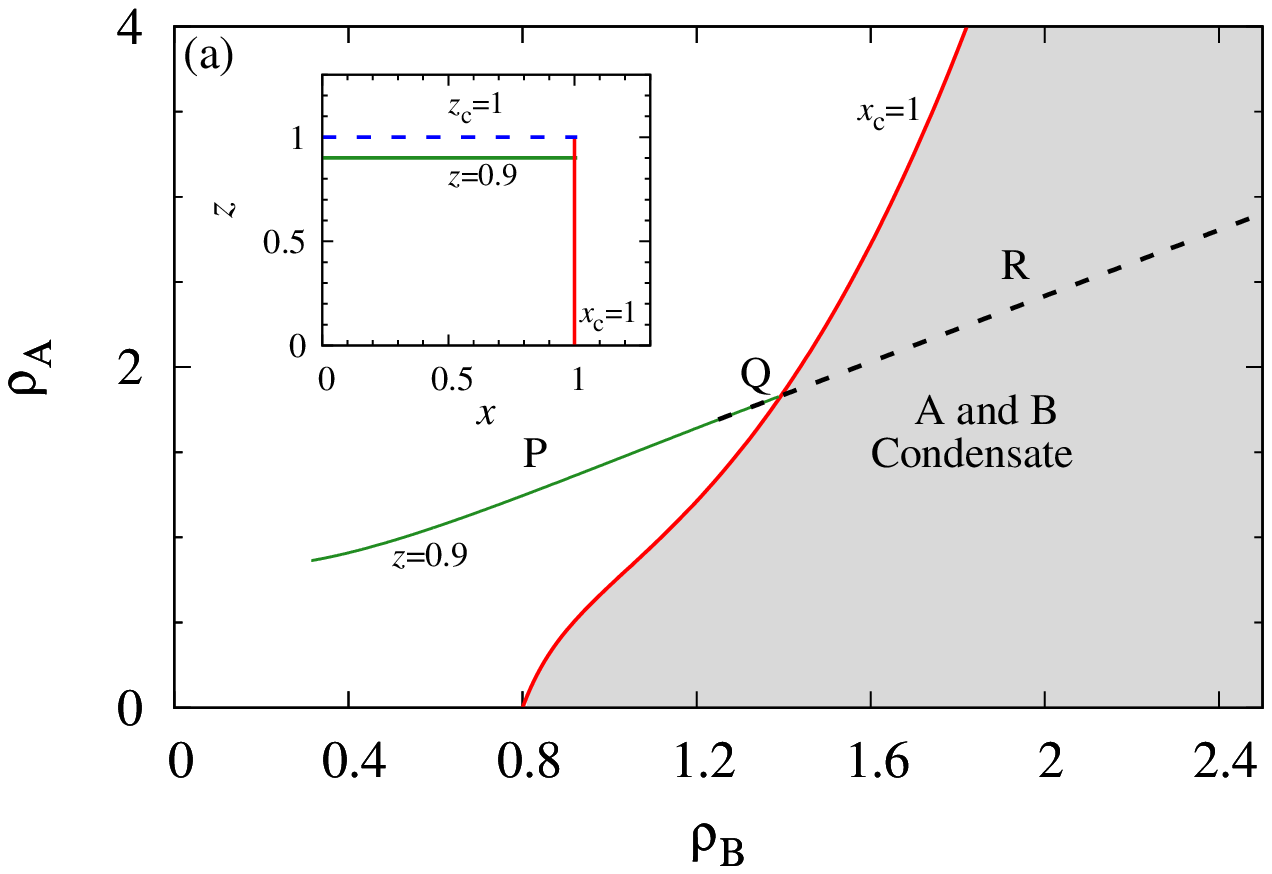}
\includegraphics[height=5.5cm]{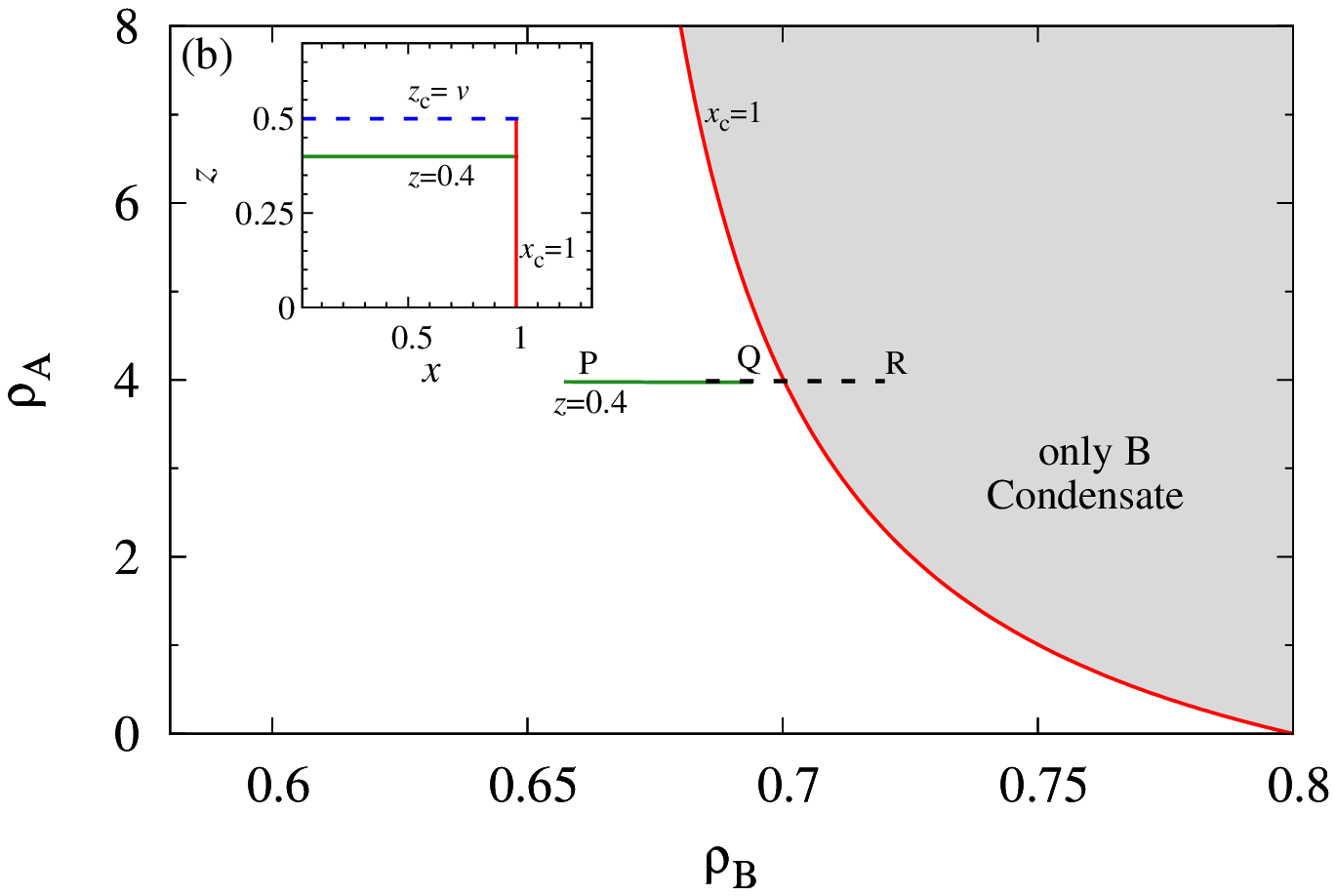}
\caption{Phase diagram in the $\rho_B$$-$$\rho_A$ plane corresponding to the critical line $x=1$ for   (a) $v=2(v>1~regime)$
and (b) $v=0.5(v<1~regime)$. In the inset we chose 
the lines $z<z_c$ and plot it in the corresponding density plane. For $v>1$, 
the slope of the line PQR is positive and it is zero for $v<1$. The background critical 
density is given by point $Q$ \cite{Gross08}.
Here $\sigma=10$ and $k=5.$ }
\label{phase_diag}
\end{figure} 

\par In absence of particles of species $A$, the critical density  
is $\rho_B^c=4/(\sigma-5)$ and thus condensation of species $B$ occurs when $\sigma>5$ \cite{misanthrope}. 
However,  when the density of species $A$ is not equal to zero ($z>0$ case), $(\rho_A^c,\rho_B^c)$ 
depends on all three parameters, $\sigma, k ~{\rm and}~ v.$ For a given value of $k$, the nature of 
condensation is found to be different for the regimes $v<1$ and $v>1$ \cite{our_model}. For $\rho_A>0$, the background 
critical densities can be found out from the grand canonical measure by using the Gro\ss kinsky theorem \cite{Gross08}. It was 
proved in \cite{Gross08} that the normal  directions   of  the critical line  in   
$\mu_x$$-$$\mu_z$  plane (here chemical potentials   are   $\mu_{x} = \ln(x)$ and  $\mu_{z} = \ln(z)$)
translates to a direction in density plane  along which  the  background density  remain invariant. 
For the model under study,  the critical line is    $x=1$   ({\it i.e.}   $\mu_x=0$)   and thus the normals are defined by 
$z=constant$. For illustration, we  take $z=0.9 (<z_c=1)$ for $v=2$  and $z=0.4 (<z_c=v)$ for $v=0.5$ in  Figure~\ref{phase_diag}  and  plot the
corresponding line  in $\rho_B$$-$$\rho_A$ plane. It  approaches the  critical point  $Q \equiv (\rho_B^c, \rho_A^c)$     
with a   slope  given by   the tangent line  PQR. The Gro\ss kinsky criteria   indicate that   the  background critical density 
along the line  QR    in  the condensate phase   is invariant  and  is given by  the point Q. When $v>1$, for  any arbitrary
density   $(\rho_B, \rho_A)$  on the   line  QR,   $\rho_B>\rho_B^c$  and     $\rho_A>\rho_A^c$;  both  species would therefore have
extra particles   which  would form a condensate. When $v<1$, the slope of the line PQR is always $zero$, regardless 
of the value of $z$ and therefore only species $B$ would form a condensate.
\begin{figure}[h]
\centering \includegraphics[height=6.5cm]{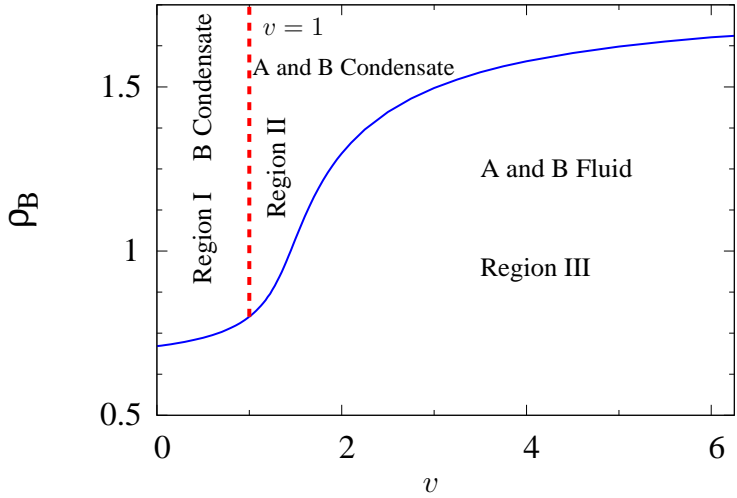}
\caption{Phase diagram of the model in $v-\rho_B$ plane for $k=5$, $\sigma=10$, and $\rho_A=1.5$. The red dashed
line, $v=1$, separates Region I where species $B$ condense and Region II where both $A$ and $B$ can condensate. The blue curve 
separates the fluid region (Region III) from  regions I and II (also see text). }
\label{reentrant_rB_v_phase_diag}
\end{figure}
\par For $v=1$, from Eq. \eqref{diff_A_species} it is clear that the dynamics of the two species gets decoupled and therefore 
species $B$ may condense with $\rho_B^c= 4/(\sigma-5)$ but
species $A$ do not have a condensate. The line $v=1$ thus separates {\it only } species $B$ condensing phase region from other regions.
In Figure~\ref{reentrant_rB_v_phase_diag}, we plot the variation of $\rho_B$ with respect to the coupling parameter $v$ for  fixed values 
of $\rho_A,~k$ and $\sigma$. Clearly, in the phase diagram, there are three regions describing different phases:
\vskip .1cm
\noindent {\bf Region~I:} Here $v<1$, species $B$ has a condensate but species $A$ remains fluid. However, from Figure~\ref{phase_diag}, it is clear 
that  for $v<1$, the liquid phase of 
$A$ at point R has a density $\rho_A^c$ and hence it remains critical.
\vskip .1cm
\noindent {\bf Region~II:} For intermediate values of $v$, both A and B would condense.
\vskip .1cm
\noindent {\bf Region~III:}  For higher values of $v$, the system remains fluid and  neither of the two species has a  
condensate. Here the fluid of both species is not critical, and is physically different from Region I, where $A$ is a critical fluid.
\vskip .1cm
\par Thus, in the phase diagram (Figure~\ref{reentrant_rB_v_phase_diag}), as we move along $v$-axis, we find 
that the system  passes successively through regions where species $A$ is fluid-condensate-fluid thereby showing a reentrant feature. For species $B$ 
the transition is from condensate to fluid phase only, and thus the reentrant behavior does not exist.

\section{Reentrant condensation transition in some single species  models}
\label{sec:other_models}
\subsection{Zero-range process}
\label{sec:zrp_reentrant}
Let us consider a zero range process \cite{evans_braz, zrp_review}, on a periodic ring with $L$ boxes or sites and $N$ 
number of particles. In ZRP, a particle can hop from a site to its neighboring site with a rate that depends only on the 
occupation number of the departure site. Each box $i$ can be occupied by any number of particles, but there is an overall 
conservation of particles since particles cannot be created or 
destroyed. If $n_i$ is the number of particles at an arbitrary site $i$, then $C$=$\{n_1,n_2, \cdots\}=\{n_i\}$ represents 
a typical configuration of the system. The particle conserving dynamics  is
defined as following: from a randomly chosen site $i$, a particle can jump asymmetrically towards its right site $i+1$ with rate $u(n_i)$
that depends only on the number of particles at $i$.  
The steady state for the ZRP  can be determined  exactly \cite{zrp_review} and can be written in the form of a product measure:
\begin{equation}
P(C)= \frac{1}{Q_{N}^L}\prod_{i=1}^L f(n_i) \delta \left(\sum_{i=1}^L n_i - N\right); ~~~f(n)=\prod_{i=1}^n \frac{1}{u(i)},
\label{eq:zrp1}
\end{equation}
where $Q_{N}^L=\sum_{\{ n_i \}} P(\{n_i\}) \delta \left(\sum_{i=1}^L n_i - N\right)$ is the canonical partition function
and the delta-function ensures that the total number of particles are conserved.
Let us make the following choice of rates
\begin{equation}
u(n)=\frac{n+ a b}{n+ a b^2}.
\label{eq:zrp2}
\end{equation}
Here $0<b<1$ and  $a>0$. Note that in the asymptotic limit, where $n \to \infty$, the above equation reads as
\begin{equation}
 u(n)= 1+ \frac{\gamma}{n} + O\left(\frac{1}{n^2} \right),
\end{equation}
where $\gamma = a b (1-b)$. It is well known \cite{zrp_review} that in ZRP, with rates having the above functional form, 
there is a condensation transition for large densities (for $\rho$ larger than a critical value $\rho_c$) if $\gamma>2$.
Thus, we expect a condensation transition here if $ab(1-b)>2$. However, the critical density $\rho_c$ depends on both $a$ and $b$ and not just 
on $\gamma=ab(1-b)$. To calculate $\rho_c$ exactly, we proceed to evaluate the partition function $Z_L(z)$ in the GCE, to obtain,
\begin{equation}
 Z_L(z)=F(z)^L,~~~~{\rm with}~~~F(z)= \sum_{n=0}^\infty z^n f(n),
\end{equation}
$z$ being the fugacity. The density in GCE is $\rho(z)= z \frac{F'(z)}{F(z)}$, and the critical density as obtained by following the same 
algebraic steps as in \cite{zrp_review} is  given by 
\begin{equation}
\rho_c= \lim_{z \rightarrow 1} \rho(z)= \lim_{z \rightarrow 1} \frac{z F^{'}(z)}{F(z)}= \frac{1+a b^2}{a b(1-b)-2}.
\label{crit_zrp}
\end{equation}

%
%

%
%

%

\begin{figure}[htbp]
\centering \includegraphics[height=6cm]{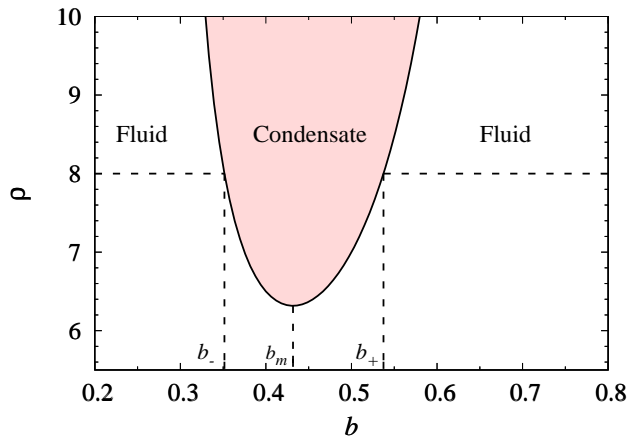}
\caption{ Critical line equation for $a=10$ having a minimum at $b_m$ is shown. As $b$ is increased, the system 
passes successively through fluid-condensate-fluid phase 
and hence it shows reentrant behavior. The reentrant transition points $b_{\mp}$ for $\rho=8$ are also shown. }
\label{reentrant_zrp}
\end{figure}

From Eq.~\eqref{crit_zrp}, it follows that for any given $a$, the critical density has a minimum at 
$b_m=\left(1+\sqrt{1+a}\right)/a$, where $\rho_c=$ $\rho_c^m=2\left(3+\sqrt{1+a}\right)/(a-8)$. Condensation can thus occur only 
if $\rho>\rho_c^m$. Further, for any $\rho>\rho_c^m$, the reentrant transition points are given by
\begin{equation}
 b_{\mp}=\frac{a\rho \mp \sqrt{a^2 \rho^2-4a(\rho+1)(2\rho+1)}}{2a(\rho+1)}.
\end{equation}
In Figure~\ref{reentrant_zrp}, we plot the  critical density $\rho_c$ as a function of $b$ for $a=10$. 
Clearly, as $b$  is increased, the system 
passes successively from fluid to condensate phase at $b_{-}$ and from condensate to fluid phase at $b_{+}$, thereby showing 
a reentrant behavior. 

\par Note that from Eq. \eqref{crit_zrp}, it is obvious that $\rho_c \sim 1/(\gamma-2)$. This indicates that reentrant transition cannot be 
observed in $\gamma-\rho$ plane as $\rho_c$ is a monotonic decreasing function of $\gamma$ (for $\gamma>2$). Thus, in this case, the reentrant 
behavior we observe is only because $\gamma$ is quadratic in $b$.

%
%

\subsection {Models with pair-factorized steady states}
\label{sec:pair_reentrant}
Let us discuss another box-particle model of one species  where the steady state has an exact pair-factorized form. 
The model \cite{evans_pair, waclaw_pair2} involves biased diffusion 
of particles on a one dimensional periodic lattice.  
Here, a particle hops from a randomly chosen site $i$ to site $i+1$ with  rate $u(n_{i-1},n_i,n_{i+1})$ that depends on the 
number of particles at site $i-1$, $i$ and $i+1$. The probability of finding the system in an arbitrary configuration 
$C=\{n_i\}$ can be expressed as a product of pair-factorized weights over consecutive sites:
\begin{equation}
P(C)=\frac{1}{Q_{N}^L} \prod_{i=1}^L g(n_i,n_{i+1}) \delta\left(\sum_{i=1}^L n_i-N\right),
\label{pair_eq:1}
\end{equation}
where $Q_{N}^L=\sum_{\{ n_i \}} P(\{n_i\}) \delta \left(\sum_{i=1}^L n_i - N\right)$ is the canonical
partition function. The $\delta$-function ensures 
that the total number of particles are conserved. It can be proved \cite{evans_pair, waclaw_pair2} that the steady state master
equation is satisfied if the rate of diffusion satisfies the following condition:
\begin{equation}
u(n_{i-1},n_i,n_{i+1})= \frac{g(n_{i-1},n_i-1) g(n_i-1,n_{i+1})} {g(n_{i-1},n_i) g(n_i,n_{i+1})}.
\label{pair_eq:2}
\end{equation}

We study a particular model with a pair factorized steady state of the form \cite{evans_pair, waclaw_pair2},

\begin{equation}
g(m,n)= \exp\left[(-J+ a U)|m-n|+ \frac{U}{2}(\delta_{m,0}+\delta_{n,0})\right],
\label{pair_eq:3}
\end{equation}

where $a\geq 0$. The corresponding hop rate  of particle diffusion  
from site $i$ to site $i+1$ obtained from Eq. \eqref{pair_eq:2} is 

\begin{equation}
u(n_{i-1},n_i,n_{i+1})= 
\begin{cases}
    \exp[2a U-2J+U \delta_{n,1}]&\text{if }n_i \leq n_{i-1},n_{i+1}\\
    \exp[2 J-2a U +U \delta_{n,1}]&\text{if }n_i > n_{i-1},n_{i+1}\\
    \exp[U \delta_{n,1}]              & \text{otherwise}.
\end{cases}
\label{pair_eq:4}
\end{equation}

\begin{figure}[htbp]
\centering \includegraphics[height=5.5cm]{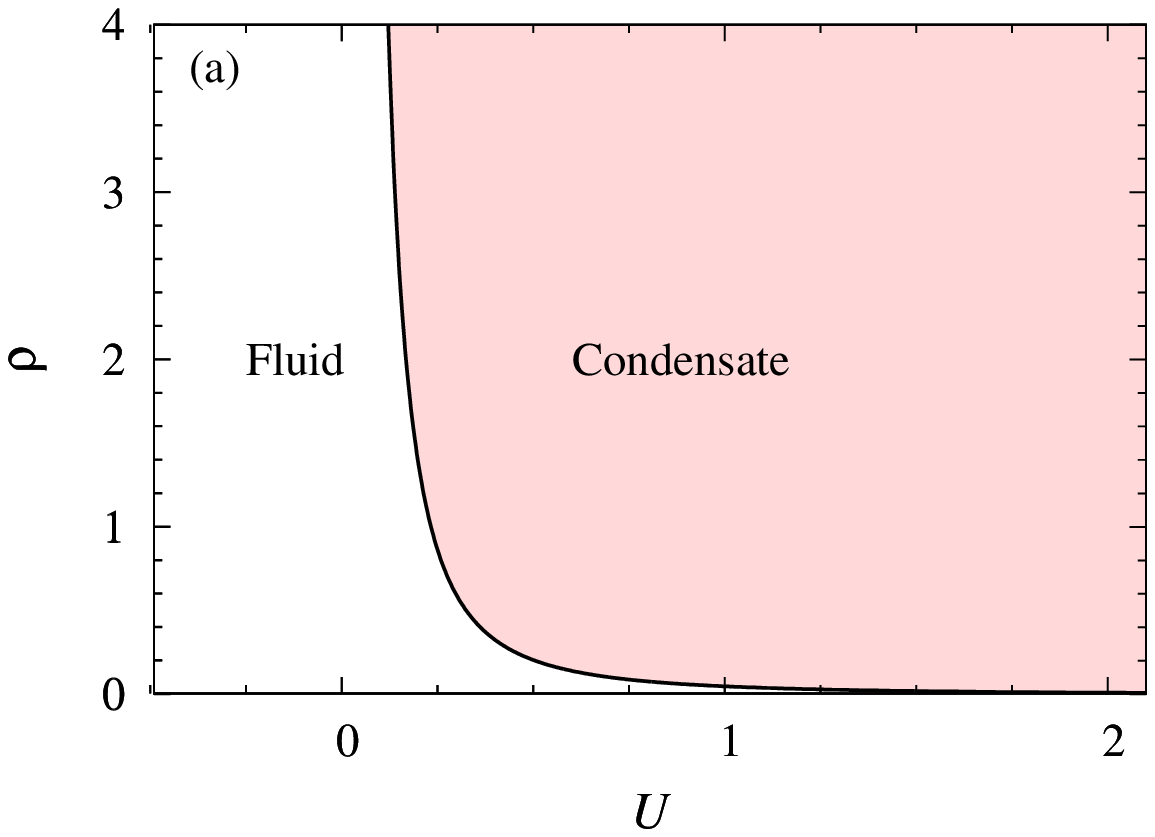} 
\includegraphics[height=5.5cm]{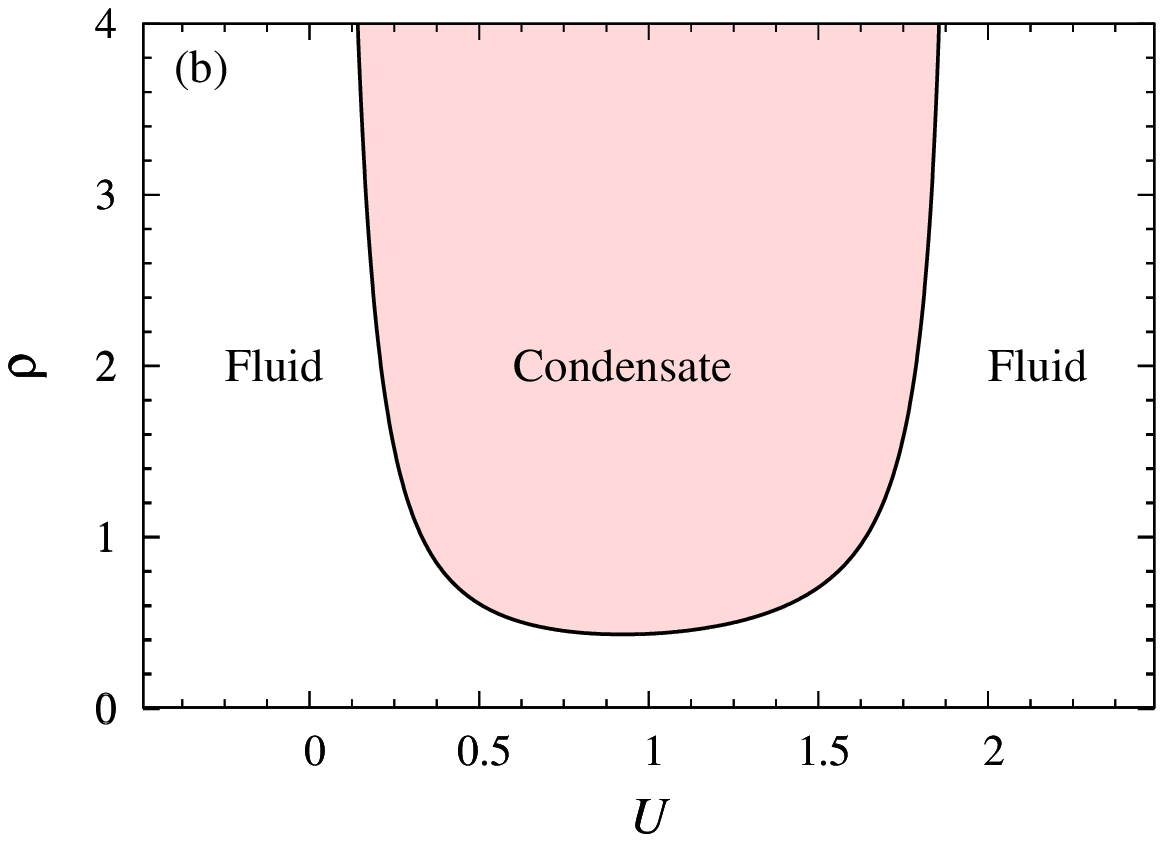}
\caption { Phase diagram for $J=2$. (a) No reentrance for $a=0$. As we move along the $U$-axis the system makes transition 
from fluid phase to condensate phase. 
(b) Reentrance for $a=1$. In this case, as $U$ is increased 
the system pass through through fluid-condensate-fluid  phase and hence the transition is reentrant in nature.}
\label{pair_figure}
\end{figure}

From Eq.~\eqref{pair_eq:4} it follows that if the number of particles at the diffusing site is smaller than the number 
of particles in the neighboring sites, the rate of diffusion is low if $J>aU$ and high if $J<aU$. Conversely,
if the number of particles at the departure site is greater than the number of particles at neighboring sites, the rate of diffusion is 
low $J<aU$ and high if $J>aU$. Thus the two factors $e^{2 J}$ and $e^{2 aU}$ compete with each other in determining
the evolution of the system. Additionally, a site can get rid of a single particle with rate $e^U$.
Note that when $a=0$, there is no competition and the model is same as 
studied  in \cite{evans_pair, waclaw_pair2}. For such a model phase transition from a fluid phase to condensate carrying phase 
do not show reentrant behavior. 
However, when 
$a>0$, we find that the system shows a reentrant behavior due to the presence of the competing terms in the interaction.
In GCE, taking $z$ as the fugacity, one can proceed with the same algebraic 
steps as in \cite{waclaw_pair2} to find that in the limit $z\rightarrow 1$, the critical density is given by 
\begin{equation}
\rho_c = \frac{e^{J_0-aU}-1}{(e^{J_0-aU}- e^{-2(J-J_0)})\times(e^{2(J-J_0)}-1)},
\label{pair_crit_den}
\end{equation}
where
\begin{equation}
J_0=U + a U -ln (e^U -1).
\label{pair_eq:6}
\end{equation}
Note that  the critical density, $\rho_c$ is finite provided that $J>J_0$. In Figure~\ref{pair_figure}
we take $J=2$ and plot the phase diagram and the critical density (Eq.~\eqref{pair_crit_den}) for two different
values of the parameter $a$. When $a=0$, the system shows a transition from fluid phase to a condensate carrying phase. 
For $a>0$,  the transition is of the nature fluid-condensate-fluid and hence shows  a reentrant behavior.
\par We would now like to point out certain differences in reentrant transitions in the models discussed above. Firstly, note that although reentrant
transition can be simply observed in some single species interacting particle systems, our model provides a clear example where reentrant transitions
can occur by tuning the interaction parameter between the two species. Secondly, in Region I of the phase diagram (Figure~\ref{reentrant_rB_v_phase_diag}), 
species $A$ always remains a critical fluid. This nature is different 
from transition in single species models discussed above where the fluid phases are never critical.

\section{Summary and conclusion}
\label{sec:summary}
In summary, we have studied the condensation transition in a box-particle system 
of two species $A$ and $B$ respectively. The rate of diffusion of the $A$ species from a given site to its next site depends 
on the number of particles of the species $B$ at that site. The dynamics of the species $B$  is a nearest neighbor exchange process, 
where by two adjacent sites can exchange particles of the species $B$ among each other. This process can 
occur only when  are two boxes participating in exchange of particles of the species $B$ are devoid of particles of 
$A$ species.  We have studied the nature of condensation transition as the interaction parameter that couples   the dynamics 
of the two species is varied and found that the condensation  in species $A$ has a reentrant feature. We have also provided two examples 
of single species driven diffusive systems which show reentrant condensation.
\par A condensation transition in a box-particle system corresponds to a phase separation transition in the
corresponding lattice model \cite{psep_criterion}. Therefore the phase separation  transition in the corresponding lattice map 
\cite{our_model} of the model studied here is also expected to show a reentrant behavior. Finally, we think that it would be interesting
to find out if reentrant  phase behavior could also be observed in a more general class of DDS, 
namely the finite-range processes which provide cluster factorised steady states \cite{finite_range}.
\section{Acknowledgements}
\label{sec:acknowledgements}
We  thank P. K. Mohanty for many useful discussions, critical comments and careful reading of the manuscript. We thankfully acknowledge the referees 
for pointing out important references and useful suggestions.


%
%

\end{document}